\documentclass[aps,prl,twocolumn,preprintnumbers,amsmath,amssymb,superscriptaddress,nofootinbib]{revtex4-1}  
\usepackage[linktocpage,pagebackref=false,hidelinks]{hyperref}

\usepackage{amsmath,setspace,amsfonts,latexsym}
\usepackage{amssymb}
\usepackage{color}
\usepackage{epsfig}
\usepackage{graphicx}
\usepackage{slashed}
\usepackage{ulem}

\definecolor{White}{rgb}{1,1,1}
\definecolor{Red}{rgb}{1,0.1,0}
\definecolor{LightYellow}{rgb}{1,1,.875}
\definecolor{SteelBlue}{rgb}{.273,.508,.703}
\definecolor{navy}{rgb}{0,0,.5}
\definecolor{LightCyan}{rgb}{.875,1,1}
\definecolor{DarkRed}{rgb}{.543,0,0}
\definecolor{HotPink}{rgb}{1,.41,.70}
\definecolor{ForestGreen}{rgb}{.13,.54,.13}
\definecolor{OliveDrab}{rgb}{.42,.55,.14}
\definecolor{MediumBlue}{rgb}{0,0,.80}
\definecolor{RoyalBlue}{rgb}{.25,.41,.88}
\definecolor{DeepSkyBlue}{rgb}{0,.746,1}
\definecolor{Brown}{rgb}{0.545,0.271,0.074}
\definecolor{Purple}{rgb}{0.637,0.285,0.641}

\def\bea{\begin{eqnarray}}
\def\eea{\end{eqnarray}}
\def\bec{\begin{center}}
\def\ec{\end{center}}

\def\beq{\begin{equation}}
\def\eeq{\end{equation}}

\newcommand\lsim{\mathrel{\rlap{\lower4pt\hbox{\hskip1pt$\sim$}}
    \raise1pt\hbox{$<$}}}
\newcommand\gsim{\mathrel{\rlap{\lower4pt\hbox{\hskip1pt$\sim$}}
    \raise1pt\hbox{$>$}}}
\def\bea{\begin{eqnarray}}
\def\eea{\end{eqnarray}}
\def\ba{\begin{array}}
\def\ea{\end{array}}
\def\bc{\begin{center}}
\def\ec{\end{center}}
\def\nn{\nonumber}

\def\met{\slashed{E}_T}

\def\beq{\begin{equation}}
\def\eeq{\end{equation}}

\newcommand{\ud}{\textrm{d}}

\begin{document}
\title{Spectral Decomposition of Missing Transverse Energy at Hadron Colliders}
\author{Kyu Jung Bae}
\email{kyujungbae@ibs.re.kr}
\affiliation{Center for Theoretical Physics of the Universe, Institute for Basic Science (IBS), Daejeon, 34051, Korea}
\author{Tae Hyun Jung}
\email{thjung0720@ibs.re.kr}
\affiliation{Center for Theoretical Physics of the Universe, Institute for Basic Science (IBS), Daejeon, 34051, Korea}
\author{Myeonghun Park}
\email{parc.seoultech@seoultech.ac.kr}
\affiliation{Center for Theoretical Physics of the Universe, Institute for Basic Science (IBS), Daejeon, 34051, Korea}
\affiliation{School of Liberal Arts, Seoul-Tech, Seoul 139-743, Korea}
\date{June 14, 2017}

\preprint{
CTPU-17-17
}

\begin{abstract}
We propose a spectral decomposition to systematically extract information of dark matter at hadron colliders.
The differential cross section of events with missing transverse energy ($\met$) can be expressed by a linear combination of basis functions.
In the case of $s$-channel mediator models for dark matter particle production, 
basis functions are identified with the differential cross sections of subprocesses of virtual mediator and visible particle production
while the coefficients of basis functions correspond to dark matter invariant mass distribution
in the manner of the K{\"a}ll{\'e}n-Lehmann spectral decomposition.
For a given $\met$ data set and mediator model, we show that one can 
differentiate a certain dark matter-mediator interaction from another through spectral decomposition. 
\end{abstract}


\maketitle
\noindent{\bf Introduction} 
Cosmological and astrophysical observations have seen strong clues of dark matter (DM) from its gravitational interaction.
For its observed thermal relic density, DM particles are believed to have non-gravitational interactions with the Standard Model (SM)
particles, for example, weakly interacting massive particles\,\cite{Lee:1977ua, Jungman:1995df}.
In order to probe such DM particles, many experiments have been conducted\,\cite{currentDD,fermi}.

Understanding DM production processes at colliders is
of great importance for the investigation of DM annihilation in the early Universe
due to the time reversal symmetry\,\cite{timeReversal}.
To identify interactions between DM and SM particles, 
many studies have utilized initial state radiation (ISR) with a missing (transverse) energy at linear colliders\,\cite{DMidentifyILC1,DMidentifyILC2} and the Large Hadron Collider (LHC)\,\cite{Barducci:2016fue,Belyaev:2016pxe}.  

Here we point out that one of the best ways to analyze DM signals at colliders is to reconstruct DM invariant mass ($m_{\chi\chi}$) distributions.
By looking at $m_{\chi\chi}$ distribution, we can extract many properties of dark sector: e.g., masses and spins of DM particles and information about the mediator(s).
In the case of a linear collider, we know the initial energy and momentum, so we are able to reconstruct $m_{\chi\chi}$ from the recoil energy,  $m_{\chi\chi}^2=(P_0-\sum_{\rm vis} P_{\rm vis})^2$ where $P_0=(E_{\rm CM}, ~\vec{0})$ is the initial four momentum and $P_{\rm vis}$ are four momenta of outgoing visible particles. 
In contrast, the $m_{\chi\chi}$ reconstruction is not available at hadron colliders due to the ignorance of initial beam-directional momentum $P_z$ of incoming partons.
Alternatively, we can utilize transverse momenta of visible particles to reconstruct a missing transeverse energy, $\slashed{E}_T=|\sum_{\rm vis}\vec{P}^T_{\rm vis}|$ where $ \vec{P}^T_{\rm vis}$'s are transverse components of three momenta of visible particles.
For this reason, previous studies of analyzing DM signals at hadron colliders had to rely on the template fitting method simulated by Monte Carlo (MC) tools which maps models to the $\met$ distribution.
However, this approach is highly model dependent.
 To cover various DM models, we need to generate corresponding MC simulations for each DM scenario.

In this letter, we propose a spectral decomposition to extract $m_{\chi\chi}$ distribution at hadron colliders from the $\met$ distribution.
Spectral decomposition has been used in various fields.
One of the most famous examples is Fourier transformation, the decomposition of a function into the linear combination of sinusoidal functions.
In a similar way, we define proper basis functions 
and decompose the $\met$ distribution into the linear combination of basis functions.
The coefficients correspond to the DM invariant mass distribution.
Note that basis functions in the $\met$ space have to be linearly independent 
but not necessarily orthogonal unlike the Fourier analysis.

\noindent{\bf Method}
For simplicity and comprehensibility, we concentrate on $s$-channel scalar mediator.
Our method is applicable to cases of $s$-channel vector mediators and we summarize short proof in the Supplemental Material.
We leave the study of the $t$-channel mediator for future work.

\begin{figure}[t] 
\begin{center}
\includegraphics[width=0.36\textwidth]{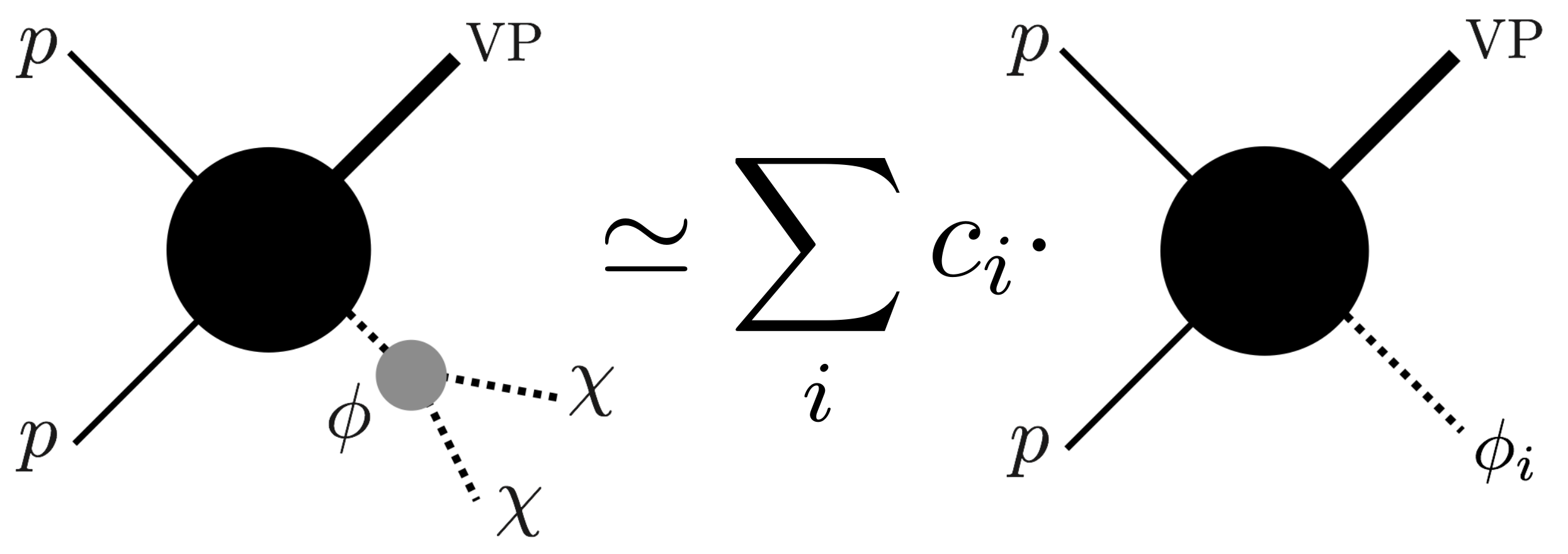}
\caption{A schematic diagram of the spectral decomposition of DM production.}
\label{diagram}
\end{center}
\end{figure}

Basis functions used in the spectral decomposition are defined by differential cross sections, the Feynman diagram of which is given in Fig.\,\ref{diagram} (right).
In Fig.\,\ref{diagram}, VP denotes associated visible particles, and $\phi$ is the physical mediator whose mass is $M_\phi$. 
In order to define basis functions, we introduce {\it virtual} mediators $\{\phi_i\}$ whose masses are assigned according to the invariant mass of dark matter particles, $m_{\phi_i}=m_{\chi\chi}^{(i)}$ for $i\in \{1,\cdots, N\}$.

With a set of basis functions, the spectral decomposition can be understood by Fig.\,\ref{diagram}; 
the differential cross section of the {\it DM production associated with VP} [Fig.\,\ref{diagram} (left)] is described by the linear combination of differential cross section of {\it the virtual mediator production} [Fig.\,\ref{diagram} (right)].
Mathematically, it is expressed as
\beq
\frac{\ud{\sigma}^{\rm exp}(X)}{\ud X}\simeq\sum_{i=1}^N c_i \underbrace{\left(\frac{1}{{\cal N}_i}\frac{\ud{\sigma}_{\phi_i}(X)}{\ud X}\right)}_{=\text{basis functions}}\,,
\label{decompositionET2}
\eeq
where $X$ is a collider observable, {\it e.g.}, $\met$ or the transverse momentum of the ISR jet.
 ${\cal N}_i$s are normalization factors, $\ud \sigma^{\rm exp}/\ud X$ is the differential cross section of physical process ($pp\to {\rm VP}+{\rm DM}$: Fig.\,\ref{diagram} (left)) and  $\ud \sigma_{\phi_i}/\ud X$ is the differential cross section of the virtual mediator production ($pp\to {\rm VP}+ \phi_i$: Fig.\,\ref{diagram} (right)).
The normalization factor ${\cal N}_i$ is given by 
\bea
{\cal N}_i=\int_{X_{\rm min}}^{X_{\rm max}}\ud X \frac{\ud{\sigma}_{\phi_i}(X)}{\ud X},
\label{normalization}
\eea
where $[X_{\rm min}$, $X_{\rm max}]$ is a range of $X$ determined by cuts. 
For applying our method to data analyses, we discretize $[X_{\rm min}$, $X_{\rm max}]$ into $\{X_\textrm{bin}\}$.

We regard the lhs of Eq.\,\eqref{decompositionET2} as experimental signal data after background subtraction and the rhs of \,\eqref{decompositionET2} as the model hypothesis.
Here, $c_i$s in Eq.\,\eqref{decompositionET2} are fitting parameters.  
One can obtain $\{c_i\}$ from the standard $\chi^2$ fitting, which minimizes
\begin{equation}
\chi^2=\sum_{X_{\rm bin}}\frac{\left({\rm Ex}({X_{\rm bin}}) -  {\rm SM}(X_{\rm bin}) - \sum_{i=1}^{N} c_i F_i(X_{\rm bin}) \right)^2}{{\rm Ex}(X_{\rm bin}) }
\label{chi2}
\end{equation}
where ${\rm Ex}(X_{\rm bin})$ is the experimental number of events in $X\in X_{\rm bin}$,
 ${\rm SM}(X_{\rm bin})$ is obtained by the SM background calculation,
 and $F_i(X_{\rm bin})$ corresponds to basis functions given by
 \bea
 F_i(X_{\rm bin})=\frac{L}{{\cal N}_i}\int_{X \in X_{\rm bin}} \ud X \frac{\ud{\sigma}_{\phi_i}(X)}{\ud X},
 \eea
 with a given integrated luminosity $L$.
 In order to obtain a unique solution from $\chi^2$ fitting, basis functions should be linearly independent. 

 If $X$ is $\met$ determined by ISR, differential distribution of $X$ 
depends on a hard scale of a parton distribution function (PDF). 
 When $\met$ is much smaller than $m_{\phi_i}$, the hard scale is mostly determined by $m_{\phi_i}$. 
 In the opposite case where $\met$ is larger than $m_{\phi_i}$, the hard scale is proportional to $\met$. 
 In other words, $m_{\phi_i}$ is the characteristic scale 
 which determines the shape of corresponding basis function.
In this regard, basis functions are linearly independent. 

In our analyses, we have numerically confirmed linear independence by examining 
\beq \footnotesize{
\min_{d_j} \Big[ \sum_{X_{\rm bin}} \Big( N_s F_i(X_{\rm bin})-\sum_{j\neq i}d_jF_j(X_{\rm bin})  \Big)^2/{\rm Ex}(X_{\rm bin}) \Big]\geq \epsilon ,}
\label{eq:linearindep}
\eeq
for all $i$ and given total number of signal events $N_s$ where $\min_{d_j}$ is minimization for parameters $d_j$.
A positive parameter $\epsilon$ is introduced to take into account statistical fluctuation.
In our analyses with seven basis functions and $S/B=1/100$, $\epsilon$ is 7.01 in 68\% confidence level.
A general discussion on the validity of this method can be found in Ref.~\cite{jung_park}.


In order to explicitly show the procedure, let us consider a DM model whose Lagrangian is written as
\bea
{\cal L}={\cal L}_{\rm SM}+\underbrace{{\cal L}_{\rm med-SM}+{\cal L}_{\rm med}}_{\rightarrow {\rm basis~functions}}+\underbrace{{\cal L}_{\rm med-DM}+{\cal L}_{\rm DM}}_{\rightarrow {\rm spectral~density}}\,,
\label{s-generalLagrangian}
\eea
where ${\cal L}_{\rm SM}$ is the SM Lagrangian, ${\cal L}_{\rm med-SM}$ (${\cal L}_{\rm med-DM}$) is the interaction Lagrangian between the mediator and SM particles (DM particles). ${\cal L}_{\rm med}$ (${\cal L}_{\rm DM}$) is the kinetic term for the mediator (DM particles) including its mass. 
${\cal L}_{\rm med-SM}+{\cal L}_{\rm med}$ affects only basis functions while 
${\cal L}_{\rm med-DM}+{\cal L}_{\rm DM}$ governs $c_i$.
The procedure of our method is described by following steps:
\begin{itemize}
\item[1.]{ fix ${\cal L}_{\rm med-SM}+{\cal L}_{\rm med}$ and calculate basis functions, }
\item[2.]{ obtain $\{c_i\}$ by applying Eq.\,\eqref{chi2} to the signal data,}
\item[3.]{ find proper ${\cal L}_{\rm med-DM}+{\cal L}_{\rm DM}$ that matches with obtained $\{c_i\}$.}
\end{itemize}
It is worth noting that one needs to 
specify ${\cal L}_{\rm med-SM}$ in order to construct basis functions.
For example, basis functions where the mediator is produced through gluon-gluon fusion is different from those where it is produced through quark and anti-quark annihilation.
However, CP charge of a mediator does not affect basis functions when the mediator is $s$-channel.
We have numerically confirmed that $\phi\,G_{\mu\nu} G^{\mu\nu}$ and $\phi\,G_{\mu\nu} \tilde{G}^{\mu\nu}$ amount to the same basis functions.\footnote{For other interactions, it can be inferred from Ref.\,\cite{Belyaev:2016pxe}.} 
The mediator models ({\it i.e.}, ${\cal L}_{\rm med-DM}$) 
can be inferred by other collider variables such as an angular correlation between jets in $jj+\met$ channel\,\cite{angular}. 
It is also possible
to concentrate on the test of mediator itself by using a visible decay mode and checking its consistency\,\cite{medvsDM}.

\noindent{\bf Spectral density} The physical meaning of $c_i$ is the  DM invariant mass ($m_{\chi\chi}$) distribution,
\bea
c_i\simeq \frac{\ud \sigma^{\rm exp}(m_{\chi\chi}^{(i)})}{\ud m_{\chi\chi}}\Delta m_{\chi\chi}^{(i)} \,. \label{METrho}
\eea
where $\Delta m_{\chi\chi}^{(i)}=(m_{\chi\chi}^{(i+1)}-m_{\chi\chi}^{(i-1)})/2$.
$c_i$ is related to the K{\"a}ll{\'e}n-Lehmann spectral density $\rho_{\phi\to\chi\chi}(m_{\chi\chi}^{(i)},M_\phi)$\,\cite{Peskin:1995ev} by
\beq 
c_i=2m_{\chi\chi}^{(i)}\, \Delta m_{\chi\chi}^{(i)}\, {\cal N}_i \, \rho_{\phi\to\chi\chi}(m_{\chi\chi}^{(i)},M_\phi)\,.
\eeq
The spectral density $\rho_{\phi\to\chi\chi}(m_{\chi\chi}^{(i)},M_\phi)$ is given by 
\beq
\rho_{\phi\to \chi\chi}(m_{\chi\chi}^{(i)},M_\phi)=\frac{1}{\pi}\big|G_\phi(m_{\chi\chi}^{(i)},M_\phi)\big|^2 m_{\chi\chi}^{(i)}\Gamma_{\phi_i\to \chi\chi}(m_{\chi\chi}^{(i)}),
\label{spectraldensity}
\eeq
where $G_\phi(m_{\chi\chi}^{(i)},M_\phi)$ is the propagator of $\phi$ with energy transfer $m_{\chi\chi}^{(i)}$ and on-shell mass $M_\phi$ and 
$\Gamma_{\phi_i \to \chi\chi}(m_{\chi\chi})$ is the decay rate of process $\phi_i\to \chi\chi$ with mass $m_{\phi_i}=m_{\chi\chi}$.
$\rho_{\phi\to\chi\chi}(m_{\chi\chi}^{(i)},M_\phi)$ does not depend on collider observables, $X$ (e.g. $p_T$, rapidity or $\met$, {\it etc.}) or cut variables.
In addition, it is independent of channels (e.g. mono-jet, mono-$Z$, etc.) and collider energy.
This feature guarantees that the spectral decomposition is valid up to detector level.
Mathematical  proofs are given in the Supplemental Material and its numerical validation is given in the next section.

Although basis functions depend on mediator models, the spectral decomposition method 
make analyses less model dependent.
Once we specify  ${\cal L}_{\rm med-SM}+{\cal L}_{\rm med}$ (step 1), 
we can obtain spectral density from the signal data  (step 2) 
and infer ${\cal L}_{\rm med-DM}+{\cal L}_{\rm DM}$ through physical insights  (step 3).

Here, we discuss some possible cases for the connection between spectral densities and the DM interactions in ${\cal L}_{\rm med-DM}+{\cal L}_{\rm DM}$.
When a mediator $\phi$ is heavier than DM threshold ($M_\phi > 2m_\chi$),  $\rho_{\phi\to \chi\chi}$ will be described by the Breit-Wigner distribution.
In the case of $M_\phi < 2m_\chi$, $\rho_{\phi\to \chi\chi}$ will be proportional to the power of DM's velocity $v_\chi$, $\rho_{\phi\to \chi\chi}\propto v_\chi^{2J_0+1}$\,. 
If $m_\chi<M_\phi<2m_\chi$ and the dominant annihilation can be $\chi\chi \to {\rm SM}$ particles through the mediator,\footnote{For $m_\chi > m_\phi$ case, the dominant process can be $\chi\chi \to \phi \phi$.}
it may be possible to infer the velocity dependence of DM annihilation process at the thermal freeze-out 
due to the time reversal symmetry\,\cite{DMidentifyILC1},
\beq
\langle\sigma_{\rm ann} v_\chi \rangle \equiv \sigma_{0} v_\chi^{2J_0}+\mathcal{O}\left(v_\chi^{2J_0+2}\right)\,,
\eeq
where it corresponds to $s$-wave ($p$-wave) if $J_0=0$ (1).
Some nontrivial spectral densities can be obtained when non-renormalizable operators \cite{Barducci:2016fue} or resonance spectrum ($M_\phi\simeq2m_\chi$) \cite{Chway:2015tma} are considered.

Compared to previous studies relying on Monte Carlo simulations, the spectral decomposition becomes more powerful when the dark sector is complicated.
For example, if there exist several DM species (heavier than $M_\phi$),  $m_{\chi\chi}$ distribution from the spectral decomposition has multi-threshold behavior.
At each threshold, we can count the power of $v_\chi$ in order to identify the interaction.
For another example, let us consider the production of two mediator particles; the one is on-shell ($M_{\phi_1}>2m_\chi$) and the other is off-shell ($M_{\phi_2}<2m_\chi$).
Our procedure will recover the Breit-Wigner resonance of $\phi_1\to \chi\chi$ process, standing on  the middle of continuum distribution from $\phi_2 \to \chi\chi$. 
Such a situation can be realized in various Higgs portal models, where both Higgs boson and singlet scalar produce DM particles through mixing. 
In addition, the spectral decomposition can be used to varify whether or not DM particles form a bound state. 
A bound state resonance will be located slightly below the DM threshold at $m_{\chi\chi}$ distribution. 
The theoretical prediction is obtained by solving non-relativistic Schr{\"o}dinger equation \cite{Strassler:1990nw},
and thus we may able to see the trace of Sommerfeld enhancement in the dark matter annihilation process\,\cite{Sommerfield}.

%
\begin{figure*}[t]
\center
\includegraphics[width=0.98\textwidth]{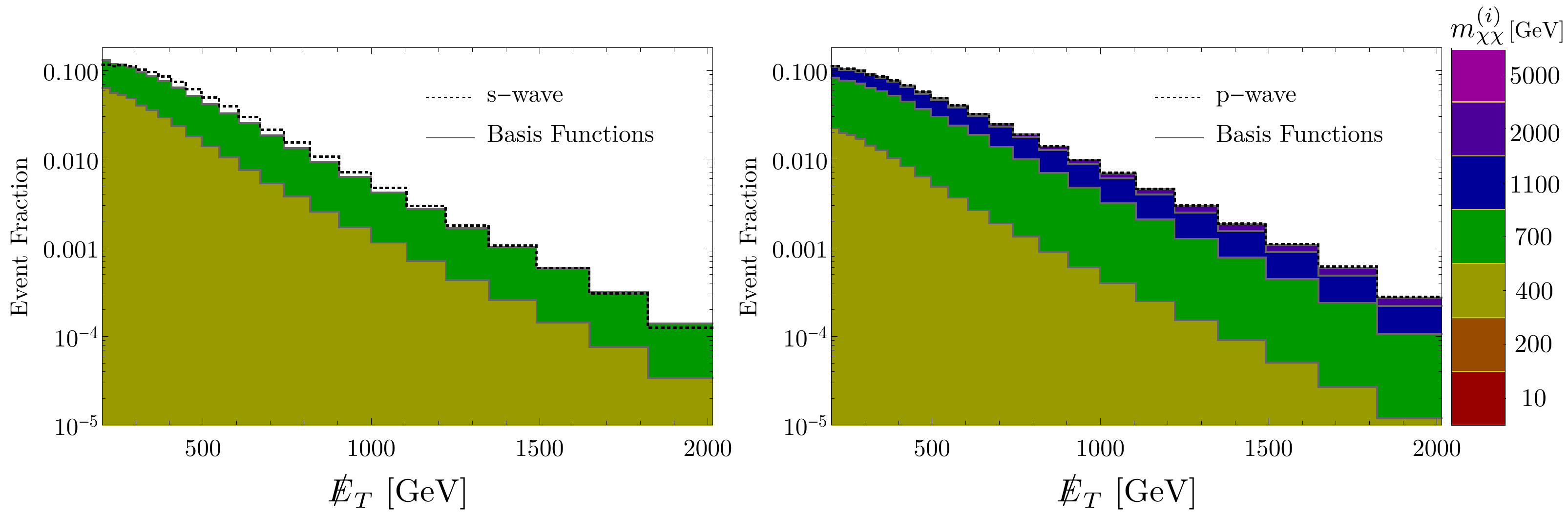}
\caption{Missing transverse energy ($\met$) distribution of mono-jet process in toy model. 
Black dotted lines correspond to the signal distributions of $s$-wave (left) and  $p$-wave (right) processes. 
Dark matter mass is set to be $200$ GeV.
Signal distributions are decomposed into basis functions (colored) whose different colors represent the value of $m_{\chi\chi}^{(i)}$, as in the  bar on the right.}
\label{decomposition_basis}
\end{figure*}
%

\noindent{\bf LHC Example.}
In order to show the detail, we provide a specific example where we set the center of mass energy to be 14 TeV and the integrated luminosity to be 3 ${\rm ab}^{-1}$.

{\it \underline{Step 1.}} Our toy model includes a real scalar mediator $\phi$ which interacts with SM through the dimension 5 operator, $\phi \,G^{\mu\nu}G_{\mu\nu}$.
We consider $X=\met$ distribution of the mono-jet process ($pp\to j\chi\chi$).
Once we fix ${\cal L}_{\rm med-SM}+{\cal L}_{\rm med}$, we can calculate basis functions ({\it i.e.}, differential cross section of ($pp\to j \phi_i$)) with a given set of $\{ m_{\chi\chi}^{(i)}\}$.
In this example, we set $\{ m_{\chi\chi}^{(i)}\}=\{ 10,\,200,\, 400,\, 700, \,1100, \,2000, \,5000\}$ in GeV units.
We take a larger gap between $m_{\chi\chi}^{(i)}$'s at higher $m_{\chi\chi}^{(i)}$ so that basis functions are 
distinguishable
in $\met$ space.
For numerical analyses, a number of tools are used; FeynRules\,2.0/MadGraph\_aMC@NLO\,\cite{MC_tools} for parton level event generation, Pythia\,8.1\,\cite{pythia8} for parton showering/hadronization and Delphes 3\,\cite{delphes}/Fastjet 3\,\cite{fastjet} for detector level event reconstruction with ATLAS detector delphes card.
Jets are reconstructed by the anti-$k_T$ algorithm with jet radius parameter 0.4.
We select events with $\met>200$ GeV and at least one jet having $p_T^{j}>100$ GeV and $|\eta^{j}|<2.5$.
We choose $\met$ bin size to be 10\% of $\met$, which corresponds to $p_T$ resolution of the jet at the LHC \cite{Khachatryan:2016kdb} and the range of $\met$ is taken from $200~{\rm GeV}$ to $2~{\rm TeV}$.

{\it \underline{Step 2.}} We decompose signal distributions into basis functions according to Eq.\,\eqref{decompositionET2} with $X=\met$. 
We generate signal distributions from pseudo-experiments (PEs).
We choose scalar dark matter (trilinear interaction with the mediator: $s$-wave) and fermionic dark matter (Yukawa interaction with the mediator: $p$-wave) for signal distributions.
A DM mass is set to be $m_\chi=200$ GeV, so the threshold is at $2m_\chi=400$ GeV.
In the Fig.\,\ref{decomposition_basis}, their $\met$ distributions are shown as black dotted lines on the top of each plot.

By following the fitting procedure explained in previous sections, we decompose signal distributions into basis functions.
In the Fig.\,\ref{decomposition_basis}, it is illustrated that basis functions (shaded by different colors) are piled into the overall distribution (black dotted line).
The area of each basis function in Fig.\,\ref{decomposition_basis} is equal to a normalized coefficient,
\bea
\hat{c}_i=c_i/\sum c_i.
\eea
which corresponds to DM invariant mass distribution.
Different colors indicate different $m_{\chi\chi}^{(i)}$s whose corresponding numbers are shown in the bar on the right-hand side.
By comparing the areas of two panels in Fig.\,\ref{decomposition_basis},  we can see that the $s$-wave process (left) tends to be more distributed around the DM threshold ($m_{\chi\chi}=$400 GeV) than the $p$-wave process (right).

\begin{figure*}[t]
\center
\includegraphics[width=0.325\textwidth]{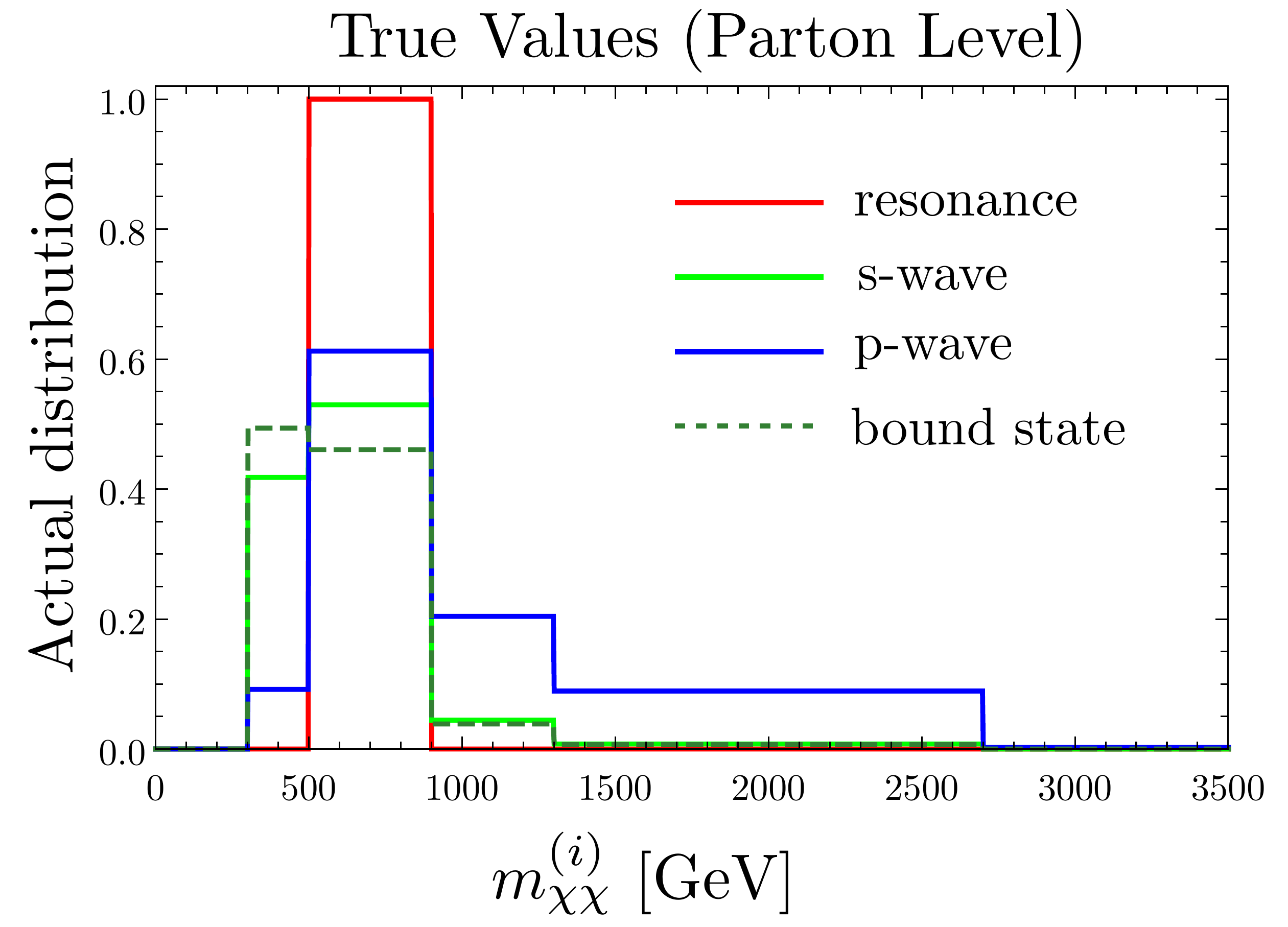}
\includegraphics[width=0.325\textwidth]{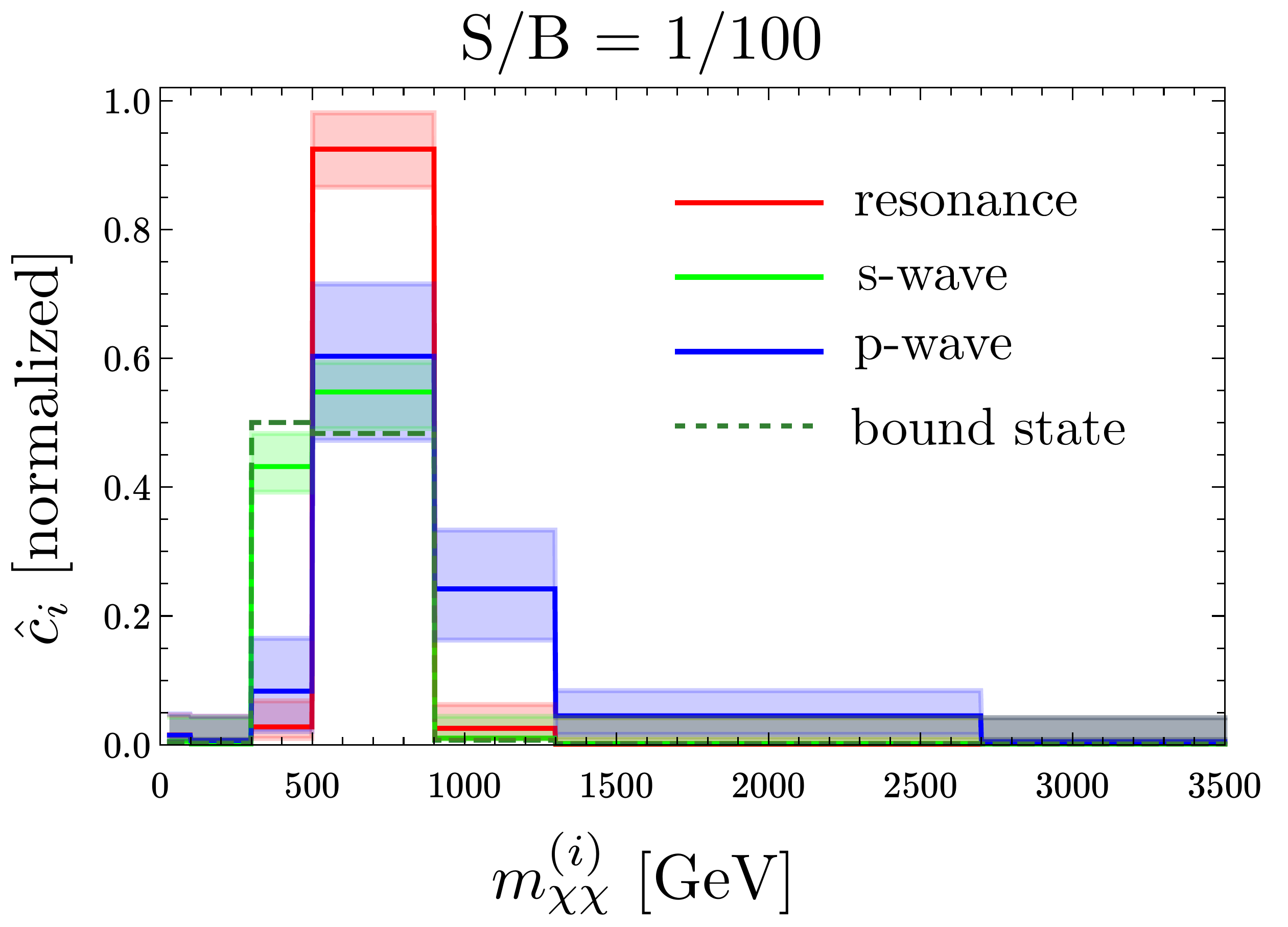}
\includegraphics[width=0.325\textwidth]{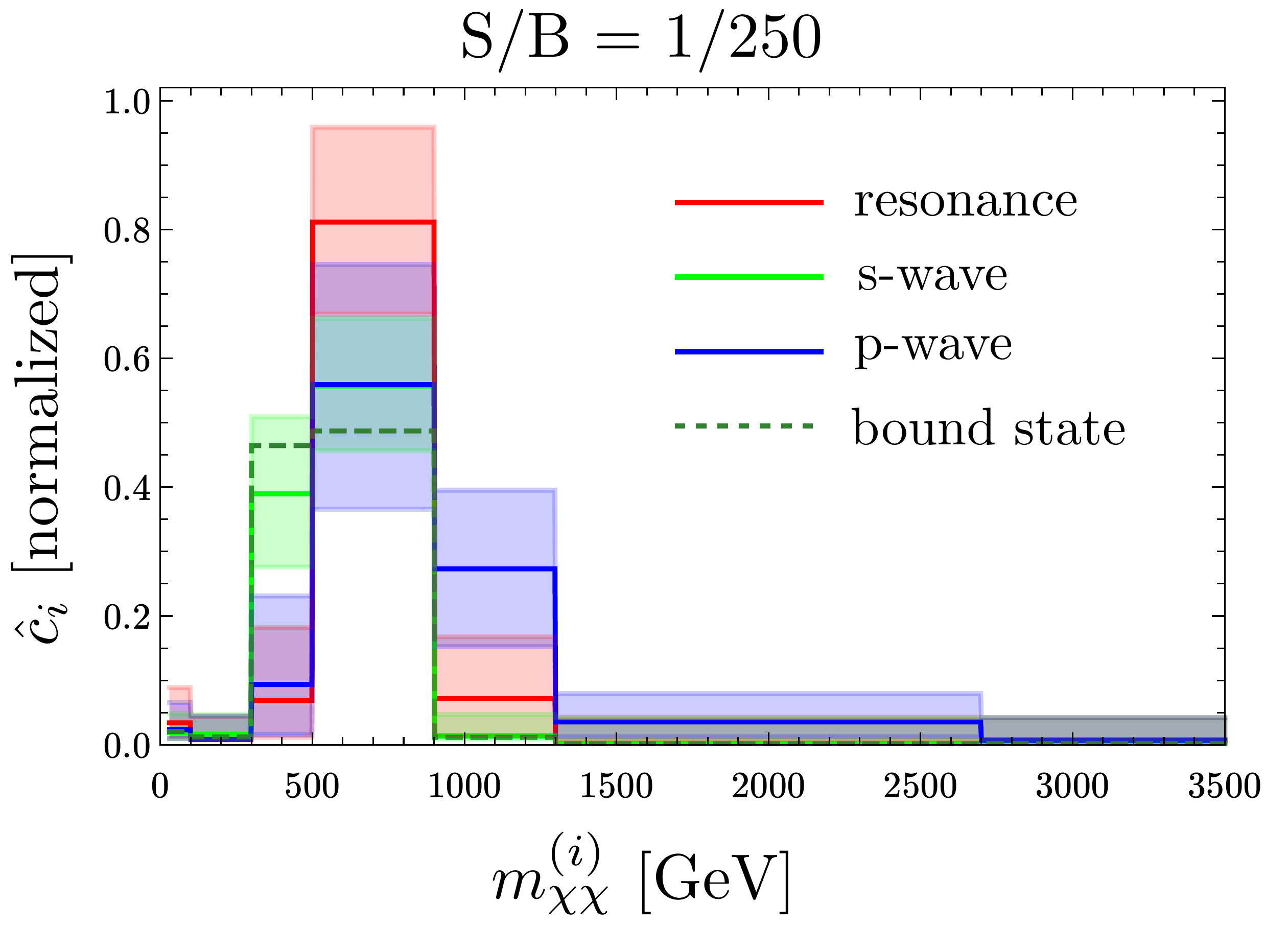}
\caption{
DM invariant mass ($m_{\chi\chi}$) distribution for DM mass $m_\chi=200$\,GeV, center of mass energy $14$\,TeV and integrated luminosity $L=3\,{\rm ab}^{-1}$.
We consider four cases: DM particles are produced in resonance decay of the $\phi$ at 700 GeV (red), in the continuum in the $s$-wave (green), in the continuum in the $p$-wave (blue) and in the continuum in the $s$-wave with bound state near threshold.
For the resonance case, $M_\phi= 700\,\textrm{GeV}\,>2m_\chi$ while for other cases, $M_\phi= 30\,\textrm{GeV}$. 
Their true distributions are given in the left panel. 
In right two panels, $\{\hat c_i\}$ obtained by our method are plotted with signal to background ratio $S/B=1/100$\,(middle) and $1/100$\,(right).
}
\label{fitting_result}
\end{figure*}

{\it \underline{Step 3}} Coefficients $\{\hat c_i\}$ show the DM invariant mass distribution from which we can infer the interaction between the mediator and DM particles.
In Fig.~\ref{fitting_result},
we show four cases of the 200 GeV DM production in $m_{\chi\chi}$ are plotted;
invisible decay of the on-shell mediator (red), 
$s$-wave through the off-shell mediator (green), 
$p$-wave through the off-shell mediator (blue) and
$s$-wave through the off-shell mediator with a DM bound state near threshold (dark green dashed).
While the true values of $\hat{c}_i$s are shown in the left panel, 
$\hat c_i$s obtained by our method  
are shown in the middle (right) panel for signal-to-background ratio $S/B=1/100$ (1/250).
On-shell production of the mediator (red) causes only one bin to be non-zero among $m_{\chi\chi}^{(i)}$s 
due to the Breit-Wigner resonance ($m_{\chi\chi}=M_\phi=700$ GeV).
For off-shell cases, coefficients are non-zero in a broad range of $m_{\chi\chi}^{(i)}$ and the first non-zero bin corresponds to the threshold, $2m_{\chi}=400$ GeV.
In the case of the $s$-wave process (green), events are more distributed near the threshold than those in the case of $p$-wave process (blue) because of $v_\chi$ dependence.
If there exists a bound state resonance (dark green dashed) 
it makes a larger value in the first nonzero bin.\footnote{Here, the bound state resonance cross section is set to be 15\% of the total cross section.
Such a case corresponds to $g_{\rm DM}^2/(4\pi)\simeq 0.35$ ($g_{\rm DM}$: gauge coupling constant in the dark sector) if DM particle is in SU(3)$_{\rm DM}$ fundamental representation.}
In middle and right panels, statistical errors are denoted by shaded bands (except for bound state case) and their discrimination power can be estimated.
While lines are separated enough to be distinguished for $S/B=1/100$, there are relatively large overlap in error bands for $S/B=1/250$.
Significances for $S/B=1/250$ are summarized in Table.\,\ref{table}.

We estimate errors in fitting $c_i$s by the following procedure.
We take the SM background of  $pp\to j+Z\,(Z \to \nu\bar{\nu})$ parton level process.
A more precise SM estimation can be found in Ref.\,\cite{LHCmonojet} which also includes $pp\to j+W\,(W \to l\nu)$ and other processes.
To make the expected number of events consistent with Ref.\,\cite{LHCmonojet}, we multiply the energy dependent $K$-factor ${\rm Max}(1,\,-\met/(400 {\rm \,GeV})+2.6)$.
The total number of signal events is fixed by setting $S/B=1/250$ and $1/100$.
For each $\met$ bin, we generate $10^4$ of pseudoexperiments (PE) by a Poissonian random number generator.
Then the fitting procedure is repeated to obtain $10^4$ different sets of $\{\hat c_i^{\rm (PE)}\}$ for each PE.
From them, we obtain probability density distribution of $\{ \hat c_i\}$ and estimate the expected value of $\{\hat c_i\}$ (solid lines) and expected 68\,\% errors (shaded bands).

From this example, we notice that a large number of signal events are required to identify the dark sector.
In Fig.\,\ref{fitting_result}, the corresponding $S/\sqrt{B}$'s are 8 (middle) and 3 (right) for $30\,{\rm fb}^{-1}$ at 13 TeV, which are almost excluded at current LHC searches.
Nevertheless, even for $S/B=1/250$ benchmark point, some cases in Fig.\,\ref{fitting_result} are distinguishable and 
we summarize the corresponding significance in Table\,\ref{table}.
\begin{table}[t]
\centering
\begin{tabular}{c||cccc}
{\footnotesize signal \textbackslash hypothesis} ~ & ~resonance  & ~$s$-wave  & ~$p$-wave  & ~bound st.   \\ \hline
resonance &                         & $1.62\sigma$  & $2.22\sigma$  & $2.15\sigma$  \\
$s$-wave & $4.30\sigma$  &                           & $4.20\sigma$ & $0.943\sigma$  \\
$p$-wave & $1.68\sigma$  & $2.21\sigma$  &                          & $2.90\sigma$  \\
bound st. & $4.48\sigma$  & $1.33\sigma$  & $4.45\sigma$ &        
\end{tabular}
\caption{ Expected significance when the signal distribution given in the first column is fitted by hypothesis given in the first row. Signal number of events is fixed by $S/B=1/250$ with $3~{\rm ab}^{-1}$ at 14 TeV.}
\label{table}
\end{table}

The resolution of $m_{\chi\chi}$ (i.e., how small bins can be) can be regarded as the distinguishability of basis functions (i.e., how large $\epsilon$ in Eq.\,\eqref{eq:linearindep} is). 
While the distinguishability is mostly affected by the statistical fluctuation in our analyses, 
a real analysis must take into account systematic uncertainties in the calculation of basis functions.
In order to precisely estimate systematic uncertainties,
full detector simulations with the best tools are required.


\noindent{\bf Conclusion and discussion.} 
The spectral decomposition allows us to extract DM invariant mass distribution even at hadron colliders.
When DM particles are produced via $s$-channel mediators, basis functions of the spectral decomposition do not depend on DM properties, while coefficients $\{ c_i\} $ (or $\rho_{\chi\to\chi\chi}$) contain detailed information of dark matter particles.

One of the most challenging issues in our approach is the requirement of a large number of signal events to identify the dark sector at hadron colliders.
Until now, the LHC has no signal excess of DM in all the mono-X channels: 
mono-jet\,\cite{LHCmonojet}, mono-photon,\cite{LHCmonophoton}, mono-Z\,\cite{LHCmonoZ} and mono-Higgs\,\cite{LHCmonoHiggs}.
The integrated luminosity is, now, about 30 ${\rm fb}^{-1}$.
This means that if we set $S/\sqrt{B} <2$ at $L=30 {\rm\,fb}^{-1}$, then it will become, at most, $S/\sqrt{B}<20$ at the end of projected high luminosity (HL) run of $L=3 {\rm\,ab}^{-1}$.
Nevertheless, it can be resolved if we combine signal data of various channels together with collider observables other than $\met$.
In addition, in the next-generation collider (e.g., 100 TeV proton-proton collider), 
better $S/B$ can be achieved in DM signals, so the spectral decomposition is useful to understand the dark sector.

\begin{acknowledgments}
\noindent{\bf Acknowlegements.}
This work is supported by the institute for basic science under the project code, IBS-R018-D1.
\end{acknowledgments}




\onecolumngrid
\section{Supplemental Material: proofs of spectral decomposition}

\subsection{Derivation of spectral decomposition}

We begin with scalar mediator and we will extend the proof to the
case where dark matter particle has a vector coupling with massive vector boson (see Sec. ``vector mediator'').
We consider a process $A+ B\to{\rm VP}+\chi\chi$ where $A$ and $B$ denote initial partons. 
$\rm VP$ stands for ``visibls particles'' (e.g. jet, photon, lepton).
For $s$-channel scalar mediator models, the scattering amplitude of $(A+B\to{\rm VP}+\chi\chi)$ is factorized as
\bea
{\cal M}_{\rm signal}={\cal M}_{A+B\to {\rm VP}+\phi} G_\phi(m_{\chi\chi}, M_\phi) {\cal M}_{\phi\to \chi\chi},
\eea
where ${\cal M}_{A+B\to {\rm VP}+\phi}$ is the amplitude of production part ($A+B\to{\rm VP}+\phi$), 
$\phi$ is a virtual mediator,
${\cal M}_{\phi\to \chi\chi}$ is amplitude of decaying part ($\phi\to\chi\chi$) and $G_\phi(m_{\chi\chi}, M_\phi)$ is the propagator of virtual mediator with a momentum transfer $m_{\chi\chi}$ and on-shell mass $M_\phi$.
By using the factorization of the amplitude, the signal cross section of ($A+B\to {\rm VP}+\chi\chi$) can be expressed by,
\begin{eqnarray}
\hat \sigma_{\rm signal}&=&\frac{1}{2s}\int \ud \Phi_{\rm VP} \ud \Phi_{\rm DM} \left|{\cal M}_{A+B\to {\rm VP}+\phi}\right|^2\left|G_\phi( p_\phi,M_{\phi})\right|^2\left|{\cal M_{\phi\to \chi\chi}}\right|^2  
(2\pi)^4\delta^{(4)}(\sum_{i\in{\rm ext}} p_i),
\eea
where $\ud \Phi_{\rm VP}= \prod_{i \in {\rm VP}}\frac{\ud ^3 \vec p_i}{2E_i(2\pi)^3}$, $\ud \Phi_{\rm DM}= \prod_{i \in {\rm DM}}\frac{\ud ^3 \vec p_i}{2E_i(2\pi)^3}$.
We denote $\hat \sigma$ ($\sigma$) by cross section calculated before (after) parton distribution function (PDF) convolution.
By inserting identity,
\bea
1=\int \ud p_\phi^0 \frac{\ud ^3 \vec p_\phi}{(2\pi)^3} (2\pi)^3 \delta^{(4)}(p_\phi-\sum_{i\in {\rm DM}} p_i)=\int \ud {m_{\chi\chi}}^2 \ud\Phi_\phi (2\pi)^4 \delta^{(4)}(p_\phi-\sum_{i\in {\rm DM}} p_i),
\eea
with $m_{\chi\chi}^2= {p_\phi^0}^2-|\vec p_\phi|^2$ and $\ud \Phi_{\rm \phi}(m_{\chi\chi})= \frac{\ud^3 \vec p_\phi}{2p_\phi^0(2\pi)^3}$, we obtain
\bea
\hat \sigma_{\rm signal}&=&\int \ud m_{\chi\chi}^2 ~ \frac{1}{2s}\int \ud \Phi_{\rm VP} \ud \Phi_{\phi} \left|{\cal M}_{A+B\to {\rm VP}+\phi}\right|^2 (2\pi)^4\delta^{(4)}(\sum_{i\in{\rm ext}} p_i)\\
&&\text{}\times|G_\phi (m_{\chi\chi},M_\phi)|^2   \frac{1}{2\pi}\int \ud \Phi_{\rm DM}   \left|{\cal M_{\phi\to \chi\chi}}\right|^2(2\pi)^4 \delta^{(4)}(p_\phi-\sum_{i\in {\rm DM}}p_i)\\
&=&\int \ud m_{\chi\chi}^2  \hat \sigma_{A+B\to {\rm VP}+\phi}(m_{\chi\chi})\cdot  \rho_{\phi\to \chi\chi}(m_{\chi\chi},M_\phi) ,
\label{decomposition}
\eea
where $\hat \sigma_{A+B\to {\rm VP}+\phi}(m_{\chi\chi})$ denotes the (virtual) mediator production cross section 
and $\rho_{\phi\to \chi\chi}(m_{\chi\chi},M_\phi)$ is a spectral density, {\it i.e.}, 
\bea
\hat \sigma_{A+B\to {\rm VP}+\phi}(m_{\chi\chi})= \frac{1}{2s}\int \ud \Phi_{\rm VP} \ud \Phi_\phi \left|{\cal M}_{A+B\to {\rm VP}+\phi}\right|^2(2\pi)^4\delta^{(4)}(\sum_{i\in{\rm ext}} p_i),
\eea
\bea
\rho_{\phi\to \chi\chi}(m_{\chi\chi},M_\phi)=|G_\phi (m_{\chi\chi},M_\phi)|^2   \frac{1}{2\pi}\int \ud \Phi_{\rm DM}   \left|{\cal M_{\phi\to \chi\chi}}\right|^2(2\pi)^4 \delta^{(4)}(p_\phi-\sum_{i\in {\rm DM}}p_i), \label{intermediate}
\eea
where mass of $\phi$ equals to the integral variable $m_{\chi\chi}$.
In Eq. \eqref{decomposition}, phase space information of visible particles is encoded solely inside $\hat \sigma_{A+B\to{\rm VP}+\phi}(m_{\chi\chi})$.
It is noteworthy that spectral density, $\rho_{\phi\to \chi\chi}(m_{\chi\chi},M_\phi)$ defined in \eqref{intermediate}  
does not depend on information of visible particles ({\it i.e.}, incoming momenta of partons and outgoing momenta of VP). 
This feature plays a crucial role in following arguments.

We take a derivative of Eq.\,\eqref{decomposition},  $\text{d}/\text{d}X$ ($X$: any function of visible particles' momenta, {\it e.g.}, $\met$, $H_T$, $p_T^{(j_1)}$, {\it etc.}),
\beq
\frac{\ud{\hat \sigma}_{\rm signal}(X)}{\ud X}=\int \ud m_{\chi\chi}^2~  \frac{\ud{\hat \sigma}_{A+B\to {\rm VP}+\phi}(m_{\chi\chi},X)}{\ud X}    \cdot   \rho_{\phi\to \chi\chi}\left(m_{\chi\chi},M_\phi\right)   \,.
\label{decompositionET}
\eeq
The differential operator ${\ud}/{\ud X}$ only acts on $\hat \sigma_{A+B\to {\rm VP}+\phi}$.
Eq.\,\eqref{decompositionET} holds after PDF convolution, {\it i.e.},
\beq
\frac{\ud{\sigma}_{\rm signal}(X)}{\ud X}=\int \ud m_{\chi\chi}^2~  \frac{\ud{\sigma}_{pp\to {\rm VP}+\phi}(m_{\chi\chi},X)}{\ud X}    \cdot   \rho_{\phi\to \chi\chi}\left(m_{\chi\chi},M_\phi\right)   \,.
\label{decompositionET}
\eeq

\subsection{Detector level}

The spectral decomposition is applicable to detector level.
Let us consider a function $F$ which transfers the parton level distribution to the detector level distribution,
\bea
F: \text{parton level distribution} &\rightarrow& \text{detector level distribution} \\
\frac{d\sigma_{\rm signal}(k_1, k_2,\cdots,k_m)}{\ud k_1 \ud k_2 \cdots \ud k_m}  &\rightarrow&  F\left(\frac{\ud\sigma_{\rm signal}(q_1, q_2,\cdots,q_n)}{\ud q_1 \ud q_2 \cdots \ud q_n}\right) ,
\eea
where $k_i$'s denote momena of visible particles at parton level and $q_i$'s denote momenta of final particles at detector level.
If we insert Eq.\,\eqref{decompositionET} into $F\left(\frac{\ud\sigma_{\rm signal}}{\ud k_1 \ud k_2 \cdots \ud k_m}\right)$, then we obtain 
\beq
F\left(\frac{\ud{\sigma}_{\rm signal}(X)}{\ud X}\right)=\int \ud m_{\chi\chi}^2~ F\left( \frac{\ud{\sigma}_{pp\to {\rm VP}+\phi}(m_{\chi\chi},X)}{\ud X}\right)    \cdot   \rho_{\phi\to \chi\chi}\left(m_{\chi\chi},M_\phi\right)   \, ,
\eeq
since $\rho_{\phi\to\chi\chi}$ does not depend on $k_i$'s.
Here, $F$ should include parton showering, hadronization, jet clustering algorithm, detector smearing effects and detector efficiency.

\subsection{Meaning of coefficients}
For each $m_{\chi\chi}$, we define the normalization function 
\bea
{\cal N}(m_{\chi\chi})=\int_{X=X_{\rm min}}^{X_{\rm max}} \ud X \frac{\ud{\sigma}_{pp\to {\rm VP}+\phi}(m_{\chi\chi},X)}{\ud X}.
\label{normalization}
\eea
From Eq. \eqref{decompositionET} and Eq.~\eqref{normalization}, we obtain
\bea
\frac{\ud{\sigma}_{\rm signal}(X)}{\ud X} &=& \int  \ud m_{\chi\chi}^2~  {\cal N}(m_{\chi\chi})  \rho_{\phi\to \chi\chi}\left(m_{\chi\chi},M_\phi\right)  \left( \frac{1}{{\cal N}(m_{\chi\chi})} \frac{\ud{\sigma}_{pp\to {\rm VP}+\phi}(m_{\chi\chi},X)}{\ud X}  \right)    \\
&=&\int  \ud m_{\chi\chi}~ 2m_{\chi\chi} {\cal N}(m_{\chi\chi})  \rho_{\phi\to \chi\chi}\left(m_{\chi\chi},M_\phi\right)  \left( \frac{1}{{\cal N}(m_{\chi\chi})} \frac{\ud{\sigma}_{pp\to {\rm VP}+\phi}(m_{\chi\chi},X)}{\ud X}  \right). 
\label{decompositionETprime}
\eea
We change integral variable $m_{\chi\chi}$ into a discrete set
$\{m_{\chi\chi}^{(1)},~m_{\chi\chi}^{(2)},~\cdots,~m_{\chi\chi}^{(N)}\}$. 
All functions of $m_{\chi\chi}$ become series with sub-index $i$ ({\it e.g.} ${\cal N}(m_{\chi\chi}^{(i)})={\cal N}_i$).
We denote $\ud {\sigma}_{pp\to {\rm VP}+\phi}(m_{\chi\chi}^{(i)},X)=\ud \sigma_{\phi_i}(X)$.
Finally, we obtain Eq.~(1) in the manuscript,
\beq
\frac{\ud{\sigma}_{\rm signal}(X)}{\ud X}\simeq\sum_{i=1}^N c_i \underbrace{\left(\frac{1}{{\cal N}_i}\frac{\ud{\sigma}_{\phi_i}(X)}{\ud X}\right)}_{=\text{basis functions}}\,,
\label{decompositionET22}
\eeq
with Eq.~(8) in the manuscript,
\beq 
c_i=2m_{\chi\chi}^{(i)}\, \Delta m_{\chi\chi}^{(i)}\, {\cal N}_i \, \rho_{\phi\to\chi\chi}(m_{\chi\chi}^{(i)},M_\phi)\,.
\eeq

On the other hand, from the Eq.\,\eqref{decompositionET}, the $m_{\chi\chi}$ distribution is given by
\bea
\frac{\ud\sigma_{\rm signal}(m_{\chi\chi})}{\ud m_{\chi\chi}}&=&\int_{X=X_{\rm min}}^{X_{\rm max}} \ud X \frac{\ud{\sigma}_{\rm signal}(X)}{\ud m_{\chi\chi} \ud X }
=\int_{X=X_{\rm min}}^{X_{\rm max}} dX 2m_{\chi\chi}  \frac{\ud{\sigma}_{pp\to {\rm VP}+\phi}(m_{\chi\chi},X)}{\ud X}    \cdot   \rho_{\phi\to \chi\chi}\left(m_{\chi\chi},M_\phi\right) \\
&=&2m_{\chi\chi} {\cal N}(m_{\chi\chi})\rho_{\phi\to\chi\chi}(m_{\chi\chi},M_\phi)=c_i/\Delta m_{\chi\chi}^{(i)}.
\eea
It proves Eq.~(7) in the manuscript,
\bea
\frac{\ud \sigma_{\rm signal}(m_{\chi\chi}^{(i)})}{\ud m_{\chi\chi}}\Delta m_{\chi\chi}^{(i)} \simeq c_i\,. \label{mxxdistribution}
\eea

\subsection{Vector mediator}
\label{sec:vec}
In this section, we show spectral decomposition for the vector mediator case.
In the unitary gauge, the tree-level propagator of the massive vector boson is given by 
\bea
G^{\rm (tree)}_\phi(k,M_\phi)^{\mu\nu} &=&\frac{-i}{k^2-M_\phi^2}\left(g^{\mu\nu}-\frac{k^\mu k^\nu}{M_\phi^2}\right) \nn\\
&=&\frac{-i}{k^2-M_\phi^2}\left(g^{\mu\nu}-\frac{k^\mu k^\nu}{k^2}\right)-\frac{-i}{M_\phi^2}\frac{k^\mu k^\nu}{k^2}
\label{vecprop}
\eea
where we separate the propagator into the transverse part ($g^{\mu\nu}-k^\mu k^\nu/k^2$) and the longitudinal part ($k^\mu k^\nu/k^2$).
This form is convenient to resum vacuum polarization tensor $\Pi_{\mu\nu}$ 
when the spectral decomposition is applied to all orders of perturbation theory.
The vaccum polarization tensor is also split into transverse and longitudinal parts,
\bea
-i\Pi_{\mu \nu}= \left(-g_{\mu\nu}+\frac{k_{ \mu}k_{ \nu}}{k^2}\right) \Pi_T+ \frac{k_{ \mu}k_{ \nu}}{k^2} \Pi_L.
\eea
After resummation of vacuum polarization, we obtain the propagator,
\bea
{G_\phi}^{\mu\nu}&=&{G_\phi^{\rm (tree)}}^{\mu\nu}+{G_\phi^{\rm (tree)}}^{\mu\rho_1}\Pi_{\rho_1 \sigma_1}{G_\phi^{\rm (tree)}}^{\sigma_1\nu}
+\cdots\\
&=&\frac{1}{k^2-M_\phi^2}\left(g^{\mu\nu}-\frac{k^\mu k^\nu}{k^2}\right)\left(1+\frac{1}{k^2-M_\phi^2}\Pi_T+\left(\frac{1}{k^2-M_\phi^2}\Pi_T\right)^2+\cdots\right) \\
&&+\frac{1}{M_\phi^2}\frac{k^\mu k^\nu}{k^2}\left(1+\frac{1}{M_\phi^2}\Pi_L+\left(\frac{1}{M_\phi^2}\Pi_L\right)^2+\cdots\right)\\
&=&G_{\rm T}(k^2, M_\phi^2)\left(g_{\mu\nu}-\frac{k_{ \mu}k_{ \nu}}{k^2}\right)-G_{\rm L}(k^2, M_\phi^2)\frac{k_{ \mu}k_{ \nu}}{k^2}
\label{vectorprop}
\eea
where $G_T=-i/(k^2-M_\phi^2-\Pi_T)$ and $G_L=i/(M_\phi^2+\Pi_L)$.

Based on above treatment of vector propagator, the scattering amplitude of ($A+B\to {\rm VP}+\chi\chi$) becomes 
\bea
{\cal M}_{\rm signal}&=&{\cal M}^\mu_{A+B\to {\rm VP}+\phi} G_\phi(m_{\chi\chi}, M_\phi)_{\mu\nu} {\cal M}_{\phi\to \chi\chi}^\nu, \\
&=&\underbrace{{\cal M}^\mu_{A+B\to {\rm VP}+\phi} G_{\rm T}(m_{\chi\chi}, M_\phi)\left(g_{\mu\nu}-\frac{k_{ \mu}k_{ \nu}}{k^2}\right) {\cal M}_{\phi\to \chi\chi}^\nu}_{=\text{transverse part}}+
\underbrace{{\cal M}^\mu_{A+B\to {\rm VP}+\phi} G_{\rm L}(m_{\chi\chi}, M_\phi)\frac{k_{ \mu}k_{ \nu}}{k^2} {\cal M}_{\phi\to \chi\chi}^\nu}_{\text{longitudinal part}} \,.  \label{vecamplitudesplit}
\eea
The transverse part of the amplitude can be expressed as
\bea
&& \left|{\cal M}_{A+B\to {\rm VP}+\phi}^\mu G_{\rm T}(k, M_{\phi})(g_{\mu\nu}-\frac{k_{ \mu} k_{ \nu}}{k^2}){\cal M^\nu_{\phi\to \chi\chi}} \right|^2\\
 &&= \left|\sum_{\lambda={\rm T,L}}{\cal M}_{A+B\to {\rm VP}+\phi}^\mu G_{\rm T}(k,M_{\phi})(\epsilon^*_{\lambda, \mu} \epsilon_{\lambda,\nu}){\cal M^\nu_{\phi\to \chi\chi}}\right|^2  \label{newpol}\\
 &&\to \frac{1}{3}\sum_{\lambda, \lambda'={\rm T,L}}\left|\left({\cal M}_{A+B\to {\rm VP}+\phi}^\mu \epsilon^*_{\lambda, \mu}\right)G_T(k,M_{\phi})\left({\cal M^\nu_{\phi\to\chi\chi}} \epsilon_{\lambda',\nu}\right)\right|^2.
\label{decorrelation}
\eea
The arrow in Eq.~\eqref{decorrelation} means that decorrelation of polarization structure is valid in calculation of cross section. 
The validity of decorrelation is shown in Refs.\,\cite{Uhlemann:2008pm}.
In the following, we summarize our version (virtual mediator case) of their proof.
As in the case of scalar mediator, cross section is written as
\bea
\hat \sigma_{\rm signal}&=&\int dm_{\chi\chi}^2  |G_T (m_{\chi\chi},M_\phi)|^2  \times\frac{1}{2s} \int d\Phi_{\rm VP} d\Phi_\phi (2\pi)^4 \delta^{(4)}(\sum_{i\in {\rm ext}}p_i) \\
&\times&\frac{1}{2\pi}\int d\Phi_{\rm DM}   (2\pi)^4 \delta^{(4)}(p_\phi-\sum_{i\in {\rm DM}}p_i)  \left|{\cal M}_{A+B\to {\rm VP}+\phi}^\mu (-g_{\mu\nu}+\frac{p_{\phi \mu} p_{\phi \nu}}{p_\phi^2}){\cal M^\nu_{\phi\to \chi\chi}}\right|^2. \label{decorrelation1}
\eea
We assume pair production of dark matter particles and we denote their four momenta by $p_1$ and $p_2$, respectively.
If we choose Gottfried-Jackson frame in which $p_\phi=(m_{\chi\chi},\vec{0})$, then $d\Phi_{\rm DM} (2\pi)^4\delta^{(4)}(p_\phi-p_1-p_2)$ can be replaced by $d\Omega_1\frac{1}{16\pi^2}\frac{|\vec p_1|}{m_{\chi\chi}}$. 
Also, the spatial component of ${\cal M}^\mu_{\phi\to\chi\chi}$ has to be proportional to $\vec p_1$ since no other vector quantities are involved in the decay process (i.e. $\vec{\cal M}_{\phi\to\chi\chi}\propto \vec p_1$). 
$\vec{\cal M}_{A+B\to {\rm VP}+\phi}^\mu $ does not depend on $\vec p_1$ and $\vec p_2$ and thus we can choose $\theta_1 =\angle(\vec{\cal M}_{A+B\to {\rm VP}+\phi}^\mu , \vec p_1)$ by rotation of the frame.
Furthermore, in the Gottfried-Jackson frame, $-g_{\mu\nu}+\frac{p_{\phi \mu} p_{\phi \nu}}{p_\phi^2}={\rm diag}(0, {\bold 1}_3)$.
Thus, eq.~(\ref{decorrelation1}) becomes
\bea
&&\frac{1}{2\pi} \int d\phi_1 \frac{1}{16\pi^2}\frac{|\vec p_1|}{m_{\chi\chi}}\int_{-1}^1 d\cos \theta_1 |\vec{\cal M}_{A+B\to {\rm VP}+\phi}^\mu |^2|\vec{\cal M}_{\phi\to \chi\chi}|^2 \cos^2\theta_1 \\
&=&\frac{1}{2\pi} \int d\phi_1 \frac{1}{16\pi^2}\frac{|\vec p_1|}{m_{\chi\chi}}\int_{-1}^1 d\cos \theta_1 \frac{1}{3}|\vec{\cal M}_{A+B\to {\rm VP}+\phi}^\mu |^2|\vec{\cal M}_{\phi\to \chi\chi}|^2. 
\eea
Finally, 
\begin{equation}
|\vec{\cal M}_{A+B\to {\rm VP}+\phi}^\mu |^2|\vec{\cal M}_{\phi\to \chi\chi}|^2
= \sum_\lambda \left({\cal M}_{A+B\to {\rm VP}+\phi}^\mu  (\epsilon^*_{\lambda, \mu} \epsilon_{\lambda,\nu}) {{\cal M}_{A+B\to {\rm VP}+\phi}^{\nu*}}\right) 
\sum_{\lambda'} \left({\cal M^\alpha}_{\phi\to\chi\chi} (\epsilon^*_{\lambda', \alpha} \epsilon_{\lambda',\beta}) {\cal M^{\beta*}}_{\phi\to\chi\chi}\right)
\end{equation}
proves that the decorrelation of polarization is valid.

On the other hand, the longitudinal part in Eq.\,\eqref{vecamplitudesplit} usually becomes zero.
If dark matter particles are scalar field, then ${\cal M}_{\phi\to \chi\chi}^\nu$ is proportional to $(p_1-p_2)^\nu$ and $p_{\phi \nu} {\cal M}_{\phi\to \chi\chi}^\nu \propto (m_1^2-m_2^2)$.
For fermionic dark matter, we define the vertex function $\Gamma_\mu$ by ${\cal M}_{\phi\to \chi\chi}^\mu \propto \bar{u_1} \Gamma^\mu v_2$.
If $\Gamma^\mu \propto \gamma^\mu$,  then $p_{\phi\mu} {\cal M}_{\phi\to \chi\chi}^\mu\propto m_1-m_2$.
If the dark matter particle has other type of interactions with vector mediator ({\it e.g.}, $\Gamma^{\mu}\propto \gamma^5\gamma^{\mu}$), the proof given in this material is not valid.


\end{document}